\documentclass[aps,twocolumn,floatfix,superscriptaddress]{revtex4}
\usepackage{epsfig}
\usepackage{subfigure}

\begin{document}

\title{Irreversible flow of vortex matter: polycrystal and amorphous phases}

\author{Paolo Moretti}
\affiliation{Departament de F\'{\i}sica Fonamental,
Facultat de F\'{\i}sica, Universitat de Barcelona, Av. Diagonal 647,
E-08028, Barcelona, Spain}

\author{M.-Carmen Miguel}
\affiliation{Departament de F\'{\i}sica Fonamental,
Facultat de F\'{\i}sica, Universitat de Barcelona, Av. Diagonal 647,
E-08028, Barcelona, Spain}

\date{\today}

\begin{abstract}
We investigate the microscopic mechanisms giving rise to plastic depinning and irreversible flow in vortex matter. The topology of the vortex array crucially determines the flow response of this system. To illustrate this claim, two limiting cases are considered: weak and strong pinning interactions. In the first case disorder is strong enough to introduce plastic effects in the vortex lattice. Diffraction patterns unveil polycrystalline lattice topology with dislocations and grain boundaries determining the electromagnetic response of the system. Filamentary flow is found to arise as a consequence of dislocation dynamics. We analize the stability of vortex lattices against the formation of grain boundaries, as well as the steady state dynamics for currents approaching the depinning critical current from above, when vortex motion is mainly localized at the grain boundaries. On the contrary, a dislocation description proves no longer adequate in the second limiting case examined. For strong pinning interactions, the vortex array appears completely amorphous and no remnant of the Abrikosov lattice order is left. Here we obtain the critical current as a function of impurity density, its scaling properties, and characterize the steady state dynamics above depinning. The plastic depinning observed in the amorphous phase is tightly connected with the emergence of channel-like flow. Our results suggest the possibility of establishing a clear distinction between two topologically disordered vortex phases: the vortex polycrystal and the amorphous vortex matter.
\end{abstract}
\maketitle

\section{Introduction}

The problem of irreversible ({\it plastic}) flow in condensed matter transport has attracted a great deal of attention over the past decades, due to its striking ubiquity. A diverse variety of physical systems exhibit features of plastic flow, including vortex matter in type II superconductors \cite{KWO-94,HEN-98}, charge density waves \cite{LEM-98}, superfluid $^4$He \cite{PEK-07} and colloidal matter \cite{PER-08}. 
Plastic transport phenomena usually emerge in driven disordered media. The collective response of such systems is well described by pinning theories \cite{NAR-92,NAT-92,CHA-00} in the limit of weak disorder, where dynamics exhibits glassy properties due to the complex balance between elasticity of the medium and disorder of the substrate. However, as soon as this balance is broken, the medium cannot behave elastically and has to deform plastically in order to accommodate strain accumulation. This is commonly observed whenever the strength of disorder overpowers lattice elasticity of the medium, or when the medium itself is too soft or dilute to adjust elastically to impurities. In that case it may either develop localized topological defects such as lattice dislocations or fall into an amorphous phase, not reminiscent of the ordered medium, if disorder is strong enough. 

Out of the several systems exhibiting this phenomenology, flux line (vortex) lattices in type II superconductors represent an ideal framework of investigation, due to their remarkable experimental accessibility and their ability to be mechanically tuned by changing the applied magnetic field $H$ \cite{BLA-94}. The vortex density is proportional to $H$ and so is the shear modulus $c_{66}$ of the vortex lattice in most cases, so that upon changing the value of the field one can control the stiffness and the response of the lattice. However, as soon as fluctuations (impurities, thermal fluctuations {\it etc.})  are able to break lattice order, mechanical properties of the now-defected vortex lattice change dramatically. 

Vortex motion is commonly induced by external current densities $J$, while collective vortex velocity is proportional to the experimentally measured voltage $V$. As a consequence, any change in the mechanical response of the vortex array is expected to affect the electrodynamic response of the underlying superconductor. In this light, anomalies in the electromagnetic behavior are usually signals of the onset of plasticity.

The breakdown of the elastic description for a vortex lattice can be observed for instance by looking at the critical current $J_c$, which is thus a plastic threshold to vortex motion.
 Experimental studies of the critical current $J_c$ include measurements of $J_c$ as a function of both the  field $H$~\cite{WOR-86,HEN-96} and the temperature~\cite{HEN-98}. All systems, however, exhibit memory effects governed by an annealing process. The system is initially in a disordered state, commonly obtained by rapidly cooling the sample from a high temperature vortex-liquid phase. This protocol, known as {\it field cooling}, ensures that the systems freezes in a metastable disordered state. As large enough currents are applied, disorder may partially heal and the critical currents measured for the healed state are always lower than those of the {\it field cooled} state. 
The annealing time is found to diverge as the critical current is approached from above \cite{HEN-96}. In fact it was shown that the two states may even coexist ~\cite{PAL-00, MAR-02}, resulting in the coexistence of a moving and a pinned phase. It is now widely accepted that such a non-annealed field-cooled state should correspond to a disordered phase. This is in agreement with the experimental observation of glassy features such as power-law distributed relaxation times~\cite{HEN-96}. 
 
The so called {\it peak effect} is yet another example of the connection between higher critical currents and incipient plasticity \cite{MAR-02}. A peak in the critical current accompanies a disordering transition at high temperature or field. Peaks in generalized resistivities however appear to be ubiquitous whenever a disordering transition is encountered, even in rather different systems such as simulated polydisperse colloids \cite{REI-08}.

Current-voltage characteristics provide a direct representation of force-velocity relations for the vortex array and allow one to disclose the nature of the plastic depinning transition as soon as $J_c$ is reached.
In agreement with the observation of coexistence phenomena, one expects the transition to be discontinuous. This is a well known experimental result \cite{JEN-88} and appears very general as it has been observed also in driven colloidal matter \cite{PER-08} as well as in charge
density waves \cite{MAE-90}. The discontinuous nature is found to increase for increasing disorder. 
 
As for the topology of the vortex array in the disordered phase, two possibilities may arise as we discussed above. (i) If the lattice is able to retain part of the topological order, it will develop dislocations in order to relax stresses. Dislocations will arrange into linear patterns that will act as grain boundaries, resulting in a vortex polycrystal.   Vortex polycrystals have been observed, after \textit{field-cooling}, in various superconducting materials such as NbMo~\cite{GRI-89,GRI-94}, NbSe$_2$~\cite{MAR-97,MAR-98,PAR-97,FAS-02}, BSSCO~\cite{LIU-94} and YBCO~\cite{HER-00}. Plastic flow properties of vortex polycrystals are determined by dislocation dynamics and have been numerically investigated in several works in the past \cite{FAL-96,CHA-02,MOR-05,MOR-09}. On the theoretical side, dislocation arguments were put forward in order to explain plastic creep \cite{NOR-00,MOR-05} and plastic depinning \cite{MOR-04,MOR-05,MOR-09}. (ii) On the other hand, if disorder is too strong, or the lattice too soft, the dislocation picture may be no longer adequate to describe the system. The emergence of completely amorphous topologies, channel-like flow at relatively low currents \cite{NOR-96,HEN-96,HEL-96}, and smectic dynamics at high drive \cite{PAR-98}, summarize the main signatures of the system's behavior in this second scenario.    

Filamentary vortex flow is often observed in simulations \cite{OLI-06,BAS-01,OLS-98,BAS-98,BAS-99}. Depending on simulation parameters, it is found to proceed either in individual well-defined channels or in several weakly coupled rivers. Such observations have also inspired a recent field theoretical formulation of the plastic depinning problem, which is supposed to provide a reliable picture of irreversible transport whenever layered dynamics occur \cite{LED-08}.

In this paper we study the mechanisms of plastic depinning and plastic flow in vortex matter by means of numerical simulations. Two limiting cases are considered. (i) The first is that of weak pinning interactions. In this case disorder is strong enough to produce plastic effects. Lattice topology is distorted by the presence of dislocation arrays, and vortex matter is arranged in a polycrystalline fashion. Critical currents for this systems have already been studied in a previous work \cite{MOR-09}. Here we focus on the stability of vortex lattices against the formation of polycrystalline ensembles, the mechanisms of healing, and the steady state dynamics for currents approaching the depinning threshold from above, when vortex motion is mainly localized at the grain boundaries. Filamentary flow is found to arise as a consequence of dislocation dynamics. (ii) The second case examined in the paper is that of strong pinning interactions. The dislocation description proves here inadequate, as the vortex array is now amorphous and no remnant of the Abrikosov lattice order is left. We study critical currents as a function of impurity densities and steady state dynamics above depinning. We find a tight connection between the nature of plastic depinning in this regime and the emergence of channel-like flow. 

The paper is organized as follows. Section \ref{numerical} provides a description of the numerical methods adopted throughout our work. Section \ref{weak} is devoted to the case of weak pinning interactions and Section \ref{strong} to the complementary results for strong pinning interactions. A discussion of the obtained results is found in Section \ref{conclusions}.

\section{Numerical methods}\label{numerical}

We simulate the zero-temperature dynamics of $N_v$ vortices in a 2D cross-section of linear size $L$. Such a length is fine-tuned in the $x$ and $y$ directions, in order to allow an ideal triangular vortex lattice to fit in the simulation box. Vortices are subject to external forces produced by the applied current density $J$ and interact with $N_p$ randomly distributed pinning centers, which reproduce the effect of oxygen vacancies or other impurities in the underlying material. Unless otherwise noted, simulations are performed in a system of size $L=36\lambda$ with periodic boundary conditions. From now on, all lengths will be expressed in units of $\lambda$. In particular, we choose a value of $\xi=0.2$ for the coherence length, corresponding to a value of $\kappa=5$ for the Ginzburg-Landau parameter. Such a  value is common in low-$T_c$ superconducting alloys, where vortex interactions are well described by Equation \ref{interaction} in the case of thin-film geometries (see below). 

The dynamics of each vortex $i$ at position ${\bf r}_i$ are described by a set of $N_v$ Langevin equations of motion of the form
\begin{equation}
\Gamma d{\bf r}_i/dt = \sum_j {\bf f}_{vv}({\bf r}_i - {\bf r}_j)+
\sum_j {\bf f}_{vp}({\bf r}_i - {\bf r}^p_j)+{\bf f}_L({\bf r}_i),
\label{eq:vf}
\end{equation}
where $\Gamma$ is an effective vortex viscosity. The first
term on the right-hand side describes
vortex mutual interactions via the long-range
force 
\begin{equation}\label{interaction}
{\bf f}_{vv}({\bf r})=AK_1\left(\frac{|{\bf r}|}{\lambda}\right)\hat{r},
\end{equation}
 where
$A=\Phi_0^2/(8\pi^2\lambda^3)$, $\Phi_0$ is the quantized flux carried
by the vortices and $K_1$
is a first order modified Bessel function \cite{degennes}.  The parameter $A$ sets the force scale throghout the simulations. The second 
contribution introduces the attraction exerted by each of the $N_p$ 
point defects on vortices. Point defects are randomly located at
positions ${\bf r}^p_i$ ($i=1,\ldots,N_p$) within the simulation box. They exert pinning forces due to a Gaussian potential of the form
\begin{equation}\label{pinning}
V({\bf r}-{\bf r}^p)=V_0\exp\left[-\left(\frac{{\bf r}-{\bf r}^p}{\xi}\right)^2\right],
\end{equation}
 whose
amplitude and standard deviation are $V_0$ and $\xi$ respectively.
If an external
current ${\bf J}({\bf r})$ is applied to the sample, it
generates a Lorentz-like force acting on the vortices 
\begin{equation}
{\bf f}_L({\bf
r})=\frac{\Phi_0}{c} {\bf J}({\bf r})\times \hat{z},
\end{equation}
 where $c$ is the speed
of light. The coupled equations of motion (\ref{eq:vf}) are
numerically solved with an adaptive step-size fifth order Runge-Kutta
algorithm, imposing periodic boundary conditions in both directions.

Transport properties in our simulations are quantified by looking at the critical current $J_c$ of the vortex array, the collective velocity of the system in the steady state $v$ (more precisely, its component along the drift direction), and the probability distribution of velocities in the steady state $p(v)$. Topology is taken care of by means of Delaunay triangulations and simulated diffraction patterns.

Simulations are performed as follows: i) an initial state of $N_v$ interacting vortices and $N_p$ fixed pinning centers is given; ii) we let the system relax until the collective velocity of the vortex array permanently reaches a value of zero; iii) starting form this configuration, different values of external driving currents $J$ are tested until the critical value $J_c$ is identified. Currents which comply with the condition $J>J_c$ identify the region of the phase diagram where, after an initial transient, the steady state is observed with non-zero average vortex velocity. 

We distinguish between two possible simulation protocols, depending on the initial state that we let relax. In the first, the initial state is characterized by a random distribution of vortices. This corresponds to a high temperature state and the subsequent relaxation to a rapid quench to zero temperature. Such a  procedure mimics what is done in {\it field-cooling} experiments (FC). In the second possible protocol, instead, we choose a perfect triangular lattice as the initial state. The experimental counterpart of this choice is a {\it zero field-cooling} experiment (ZFC) in which the vortex lattice has been built after the temperature quench.

As mentioned above, our results are organized in two main sections. In Section \ref{weak} we consider the case of weak pinning interactions, choosing a value of $V_0=0.01$ for the typical depth of the pinning potential wells, as introduced in Equation \ref{pinning}. In Section \ref{strong} we deal with strong pinning interactions, corresponding to $V_0=1$, two orders of magnitude larger than in the previous case. In both cases, the strength of the pinning potential  is measured in units of the force scale $A$ times the lengthscale $\lambda$.

\section{Weak pinning interactions: vortex polycrystal}\label{weak}
\begin{figure}
\subfigure[]{\epsfig{file=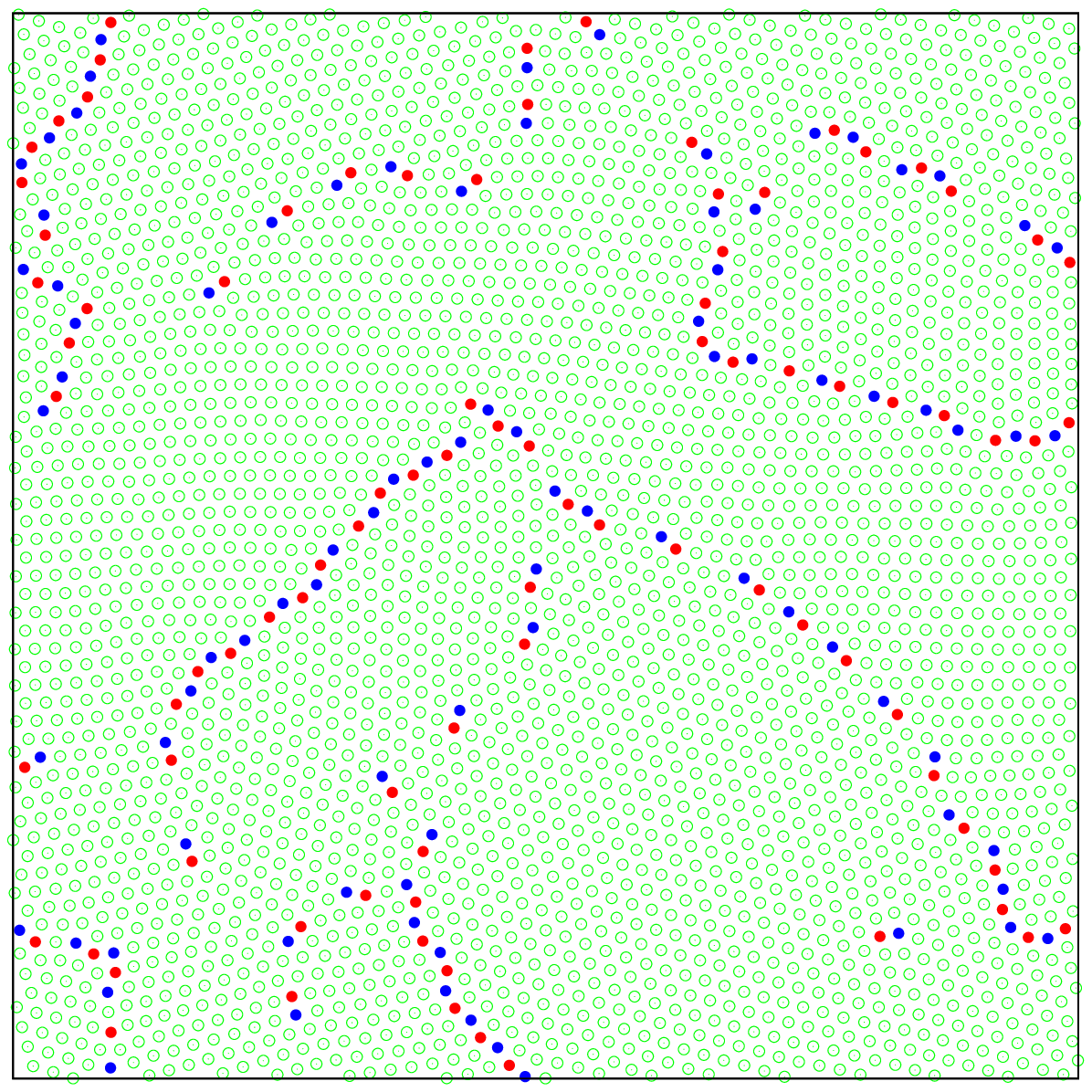,width=4cm,clip=}}
\hspace{0.2cm}\subfigure[]{\epsfig{file=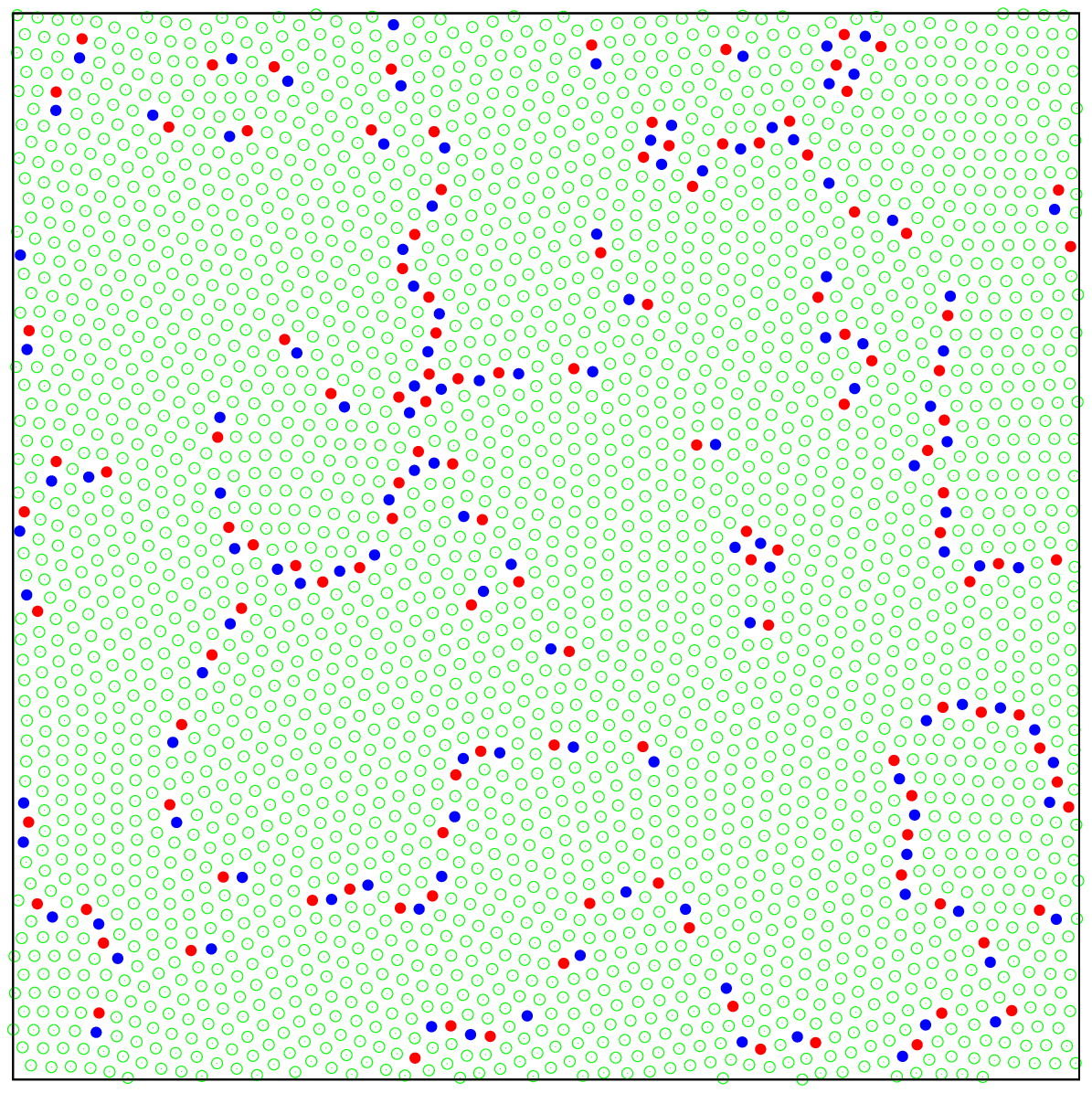,width=4cm,clip=}}
\subfigure[]{\epsfig{file=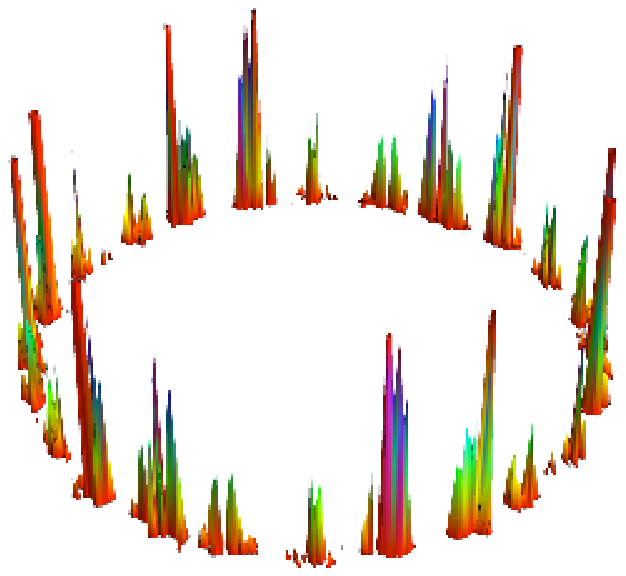,width=4cm,clip=}}
\hspace{0.2cm}\subfigure[]{\epsfig{file=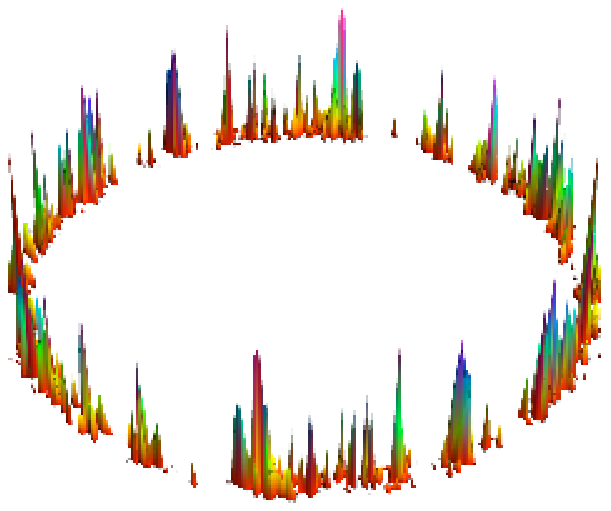,width=4cm,clip=}}
\caption{Real space images and simulated diffraction patterns for the vortex polycrystal. The case of $N_v=3120$ vortices was considered, in interaction with $N_p=1032$ pinning centers (a and c) and $N_p=8256$ pinning centers (b and d) respectively. In real space images green, red and blue circles are $6-$, $5-$ and $7-$coordinated vortices respectively. A red-blue pair corresponds to a  dislocation. A dislocation wall acts as a grain boundary, separating grains of different orientations. 
Diffraction patterns were simulated by calculating the structure factor in the first Brillouin zone. The peak in the origin was removed and lower values were cut off in order to enhance clarity. Scale factors are equal in the two plots, so that intensities in the two realizations can be compared.   The patterns reproduce multiple rotated replicas of the six-peak structure observed for perfect vortex crystals, thus signaling the emergence of polycrystalline order. For low $N_p$ grains are larger (as peaks are more intense) and typically less in number, while for large $N_p$ the grain structure becomes more complex.}
\label{fig_wtopology}
\end{figure}

In this case, starting form a random vortex configuration (FC protocol), the system relaxes into a vortex polycrystal. The system tries to recover the order of a perfect lattice, however impurities counteract this process and the sample freezes into a metastable multi-grain configuration, as in Figure \ref{fig_wtopology}. Lattice order is broken by topological defects such as dislocations, which have arranged themselves into grain boundaries. When a current above $J_c$ is driven through the sample, the vortex polycrystal depins. Depinning is nucleated at the grain boundaries and extends to the rest of the lattice as soon as large enough currents are applied \cite{MOR-09}. 

A detailed study of the topology of the relaxed state and its relationship with the critical current is reported in a previous publication \cite{MOR-09}. For the sake of completeness, we recall that the number of impurities $N_p$ affects the relaxation process. By varying $N_p$ over a range between $64$ and $10320$, we observed that all relaxed systems exhibit grain structure. However typical grain sizes are found to decrease while increasing $N_p$ only below a limiting value (roughly $N_p=4128$). Further increases of $N_p$ do not produce any appreciable drop in grain sizes. We found a striking correspondence between this behavior and the increase of the critical currents with $N_p$. In particular we found that for low values of $N_p$, where grain sizes decrease rapidly, also the critical currents increase rapidly (linearly in $N_p$). For large values of $N_p$ instead, critical currents grow with $N_p$ according to a slower square-root law. We were able to explain the behavior in terms of grain boundary pinning phenomena. Details can be found in Reference \cite{MOR-09}.

Whenever the initial condition is that of a perfect vortex lattice, as in zero-field cooling experiments (ZFC), the system always relaxes into a distorted but not defected lattice. The response to a current in such a system is the one predicted for elastic depinning \cite{CHA-00}. Vortex activity is widespread over the sample and it occurs through the propagation of avalanches. The critical current is lower than in the case of a vortex polycrystal and is proportional to the density of defects \cite{MOR-09}.

\subsection{Critical currents and stability of the vortex lattice}

Our aim is now to assess the stability of the perfect vortex lattice state with respect to the formation of a vortex polycrystal. To this end, we compared critical current characteristics of vortex lattices and vortex polycrystals for different vortex densities. The number of vortices in a fix-sized sample is proportional to the applied magnetic field H. As also the shear modulus of the lattice is proportional to the magnetic field, one can actually tune the mechanical properties of a lattice by changing the vortex density. In general one can expect that even for a defected lattice, the stiffness of the vortex array increases as the flux density is increased. Here we show that increased stiffness also produces increased stability of the vortex lattice against the formation of a vortex polycrystal. 

\begin{figure}
\epsfig{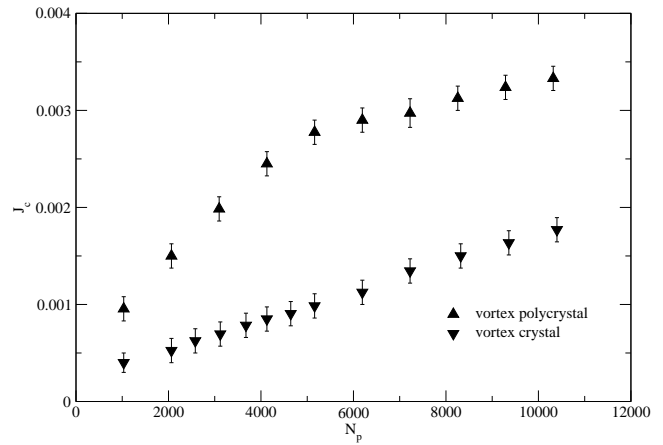}
\caption{Critical current as a function of the number of defects for a vortex polycrystal and the corresponding perfect vortex lattice in the case of large vortex densities ($N_v=3120$). The two states are well separated and no crossing in found in the examined region. Notice that a slightly modified version of this figure was first included in a previous publication (Ref. \cite{MOR-09}. It is reproduced here for the sake of completeness, as it serves a different purpose.}
\label{fig_wjcrit3}
\end{figure}

We start considering the case of high vortex densities ($N_v=3120$ vortices in the simulation box), corresponding to the highest vortex lattice stiffness considered in the simulations. We should however emphasize that such a density is well below the high-field limit where the disordering transition commonly known as {\it peak effect} takes place. In Figure \ref{fig_wjcrit3} the critical current as a function of the number of defects $N_p$ is plotted, both for the vortex polycrystal and for the vortex lattice. We observe that the two curves remain clearly separated for the whole range of pinning densities considered. Indeed, {\it exactly at depinning} the two states, produced by different sample histories, depin through different phenomena. According to the picture given above, motion in the vortex polycrystal is activated through grain-boundary depinning and plastic flow, while the perfect lattice depins elastically by releasing widespread avalanches.  
\begin{figure}
\epsfig{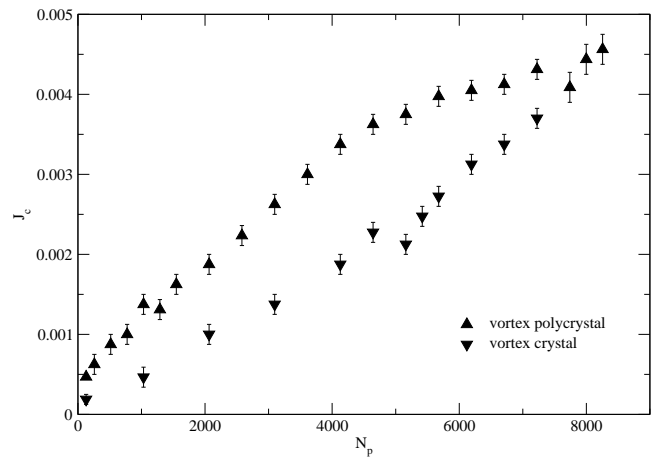}
\caption{Critical current as a function of the number of defects for a vortex polycrystal and the corresponding perfect vortex lattice in the case of intermediate vortex densities ($N_v=2016$). The two curves  eventually merge meaning that the vortex crystal is unstable against the proliferation of topological defects and the formation of a polycrystal.}
\label{fig_wjcrit2}
\end{figure}
\begin{figure}
\epsfig{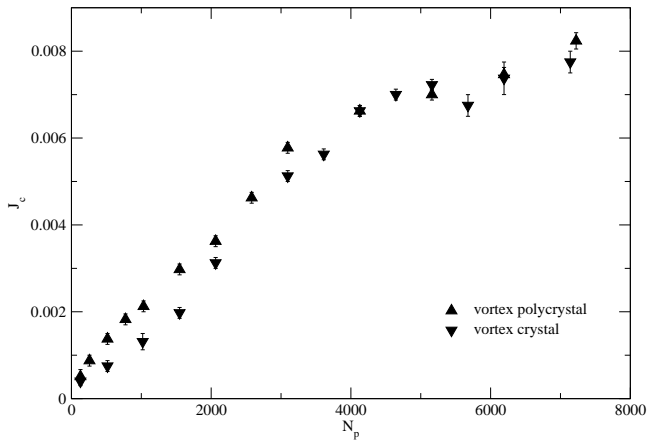}
\caption{Critical current as a function of the number of defects for a vortex polycrystal and the corresponding perfect vortex lattice in the case of low vortex densities ($N_v=1020$). The two curves almost coincide, except in the case of high purity (lowest $N_p$).}
\label{fig_wjcrit1}
\end{figure}

The fact that the two topological states are not equivalent seems obvious at first sight, but that picture changes radically as soon as the vortex density is decreased. In Figure \ref{fig_wjcrit2} we report the same $J_c$ characteristic as above, but in the case of intermediate lattice stiffness ($N_v=2016$). The separation between vortex lattice and vortex polycrystal is less evident in this case, and for high defect densities the two states are equivalent {\it at depinning} in spite of the different sample histories. By looking at Delaunay triangulations of the system, one can observe that, unlike the case of high vortex densities, here the vortex crystal is unstable against the nucleation of dislocations. The proliferation of topological defects occurs for currents just above depinning whenever the impurity density is high enough. The emergence of a dislocation network forces the perfect lattice into a state that is equivalent to a depinning polycrsytal and this is why the two current characteristics eventually merge. This phenomenon appears as a {\it re-entrant} disordered state in the dynamic phase diagram of the vortex lattice for low field. Here we provided an interpretation of such a phenomenon in terms of instability against the formation of a vortex polycrystal and showed that for intermediate vortex densities it occurs only for large enough impurity densities (high $N_p$).  

Finally we consider the case of low vortex densities ($N_v=1020$), reported in figure \ref{fig_wjcrit1}. This case is emblematic of lattice instability, as the curve for the vortex single crystal state and the curve for the vortex polycrystal state merge even at relatively low impurity densities. This means the {\it at depinning} soft vortex crystals are {\it almost always} unstable against the formation of dislocation assemblies, unless samples exhibit a high degree of purity.  

\subsection{Nature of the steady state}

So far we have focused on the response of the system at depinning, however we have not provided a description of the steady state just above depinning. It is known that upon applying large enough currents a defected vortex lattice might either heal its topological defects and collectively move as a flowing crystal or, if disorder is strong enough, or the lattice is particularly soft, it can remain topologically disordered and flow in a channel-like fashion, exhibiting smectic diffraction patterns \cite{PAR-98}. These considerations apply in the case of high driving currents and in a recent numerical study such mechanisms were detailed in the case of high AC drive (Ref. \cite{REI-06}). Here instead, we are interested in the behaviour of the system for currents close to depinning. 

\begin{figure}
\epsfig{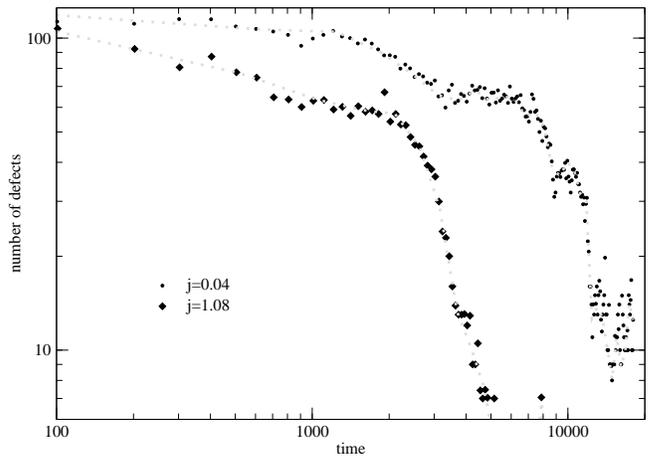}
\caption{Number of topological defects (dislocations) as a function of simulation time for high vortex density. The case of a vortex polycrystal is considered, with $N_v=3120$. Two different curves are shown, for different values of the reduced current $j=J/J_c-1$. Dotted gray lines are drown as a guide to the eye. }
\label{fig_w3defects}
\end{figure}

We can show that in the case of large vortex densities ($N_v=3120$), a vortex polycrystal always heals into an almost perfect vortex lattice, even for currents right above depinning. Such a result is reported in Figure \ref{fig_w3defects}, where the time evolution of the number of defects in a single run is recorded for different values of the reduced current $j=J/J_c-1$. For the largest current considered $j\sim 1$, the annihilation of dislocations takes place in two distinct steps: first a slow process, almost a power-law in time, then an abrupt drop in the number of dislocations, so that only a few pairs survive. Getting closer to depinning ($j\approx 0$) instead, the process takes a much larger time, as experimentally observed in Ref. \cite{HEN-96}, and goes through a series of steps in the number of defects, which suggest that the systems gets momentarily frozen in intermediate metastable states. Every time a step is overtaken, the collective velocity of the vortex assembly experiences a jump, as a decrease in the number of defects usually implies an increase in the mobility of the vortex ensemble. 

\begin{figure}
\epsfig{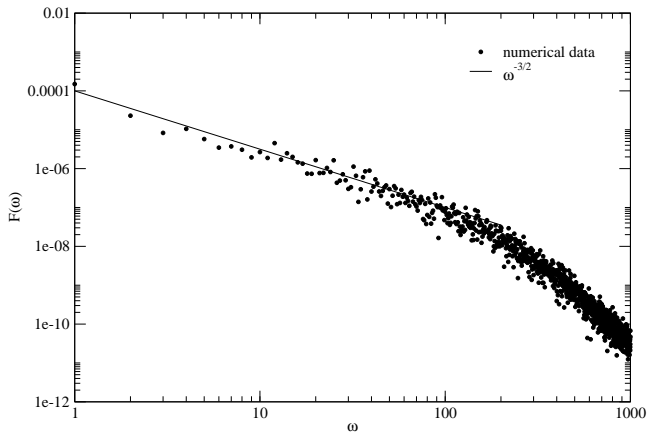}
\caption{Power spectrum of the collective vortex velocity in the drift direction at the steady state just above depinning for low vortex densities. In order to obtain a more accurate Fast Fourier Transform, we considered the case of $N_v=4032$ vortices and $N_p=16512$ pinning centers in a simulation box of $L=72\lambda$, corresponding to the case of low to intermediate vortex densities in the usual $L=36\lambda$ box. Frequencies $\omega$ are measured in units of $(t_1-t_0)^{-1}$. No washboard peak is observed. One finds a power-law decay instead, reminiscent of the $1/f$ noise associated with dislocation dynamics. The straight line is drawn for comparison to a $\omega^{-3/2}$ law. The shape of the cut-off in the power-law behavior is most probably due to the nature of the FFT routine.}
\label{fig_powerspec}
\end{figure}

In the case of low to intermediate vortex densities, however, one expects the above behavior to show substantial differences. As a matter of fact, we have already shown that even a perfect vortex lattice is in this case unstable against the proliferation of dislocations, so one can predict that a vortex polycrystal will not heal for currents just above depinning. Indeed, if we plotted the time evolution of defects as we did for high vortex densities, the curve would exhibit large fluctuations but also a constant average value in time and no decay, not even after very large waiting times. Instead of the time evolution of defects, we report the power spectrum of vortex velocities in Figure \ref{fig_powerspec}, defined as
\begin{equation}
F(\omega)=\left| \int_{t_0}^{t_1}\sum_k v_k(t)e^{i\omega t}dt \right|^2.
\end{equation}   
If the order of a perfect lattice were recovered at least to some extent, $F(\omega)$ would show a peaked structure at the so-called washboard frequency.    
However in this case we observe no peak that is reminiscent of the perfect lattice, meaning that no healing is produced in the steady state near depinning for intermediate/low vortex densities. In fact, the power spectrum shown in Figure \ref{fig_powerspec} has the form of a power-law decaying signal, commonly known as {\it broadband voltage noise} in experiments \cite{TOG-00}. This is a well established result, however here we can relate it directly to dislocation dynamics. As we know, such behavior is due to the fact that the system cannot annihilate dislocations (more precisely that it creates and annihilates dislocations at the same average rate over time). Collective dislocation dynamics is in fact responsible for the so called $1/f$ noise in plastic deformation of crystalline materials, which consists in a power spectrum that decays according to a $\omega^{-1.5}$ law \cite{LAU-06,MIG-08}, in good agreement with our results. As a consequence, the observation of broadband voltage noise appears to be again a consequence of complex dislocation dynamics in this regime. 

\begin{figure}
\epsfig{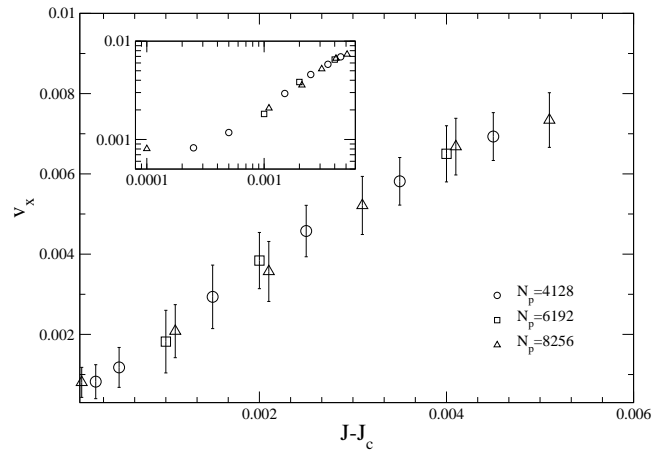}
\caption{Steady-state collective vortex velocity in the drift direction for $N_v=1020$ vortices as a function of the distance to the threshold. Different values of the defect density are considered. Inset: logarithmic plot of the same data.}
\label{fig_wvcrit}
\end{figure}

We conclude that the phenomenology of the steady state near depinning is much richer in the case of low vortex densities, as no healing is encountered and dislocation dynamics heavily affects the response of the system. It is known that in this case the vortex drift velocity at the steady state approaches zero as the applied current decreases towards $J_c$ in a weakly discontinuous fashion and this effect is found to increase in the case of larger pinning forces \cite{JEN-88,FAL-96} or alternatively  softer vortex arrays  \cite{JEN-88}. In our case however pinning forces are fixed and we are able to tune the pinning density instead. The simulated steady-state vortex velocities for $N_v=1020$ are reported in Figure \ref{fig_wvcrit}.
We found that, as expected, the transition is indeed weakly discontinuous. The curves collapse as they are plotted as a function of the distance to the threshold current and moderate variations of the defect concentration do not heavily affect the nature of the transition in this regime, where dislocation collective dynamics is supposed to control vortex flow. Of course, one cannot exclude that in principle for higher impurity densities, if the vortex ensemble developed amorphous topology, a radically different velocity-force relation could be observed. 

\begin{figure}
\epsfig{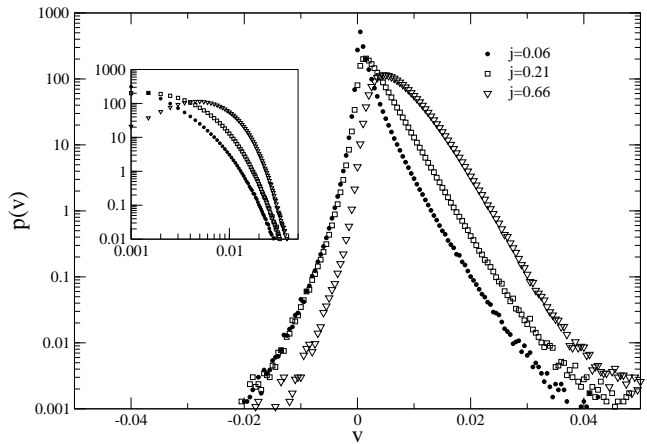}
\caption{Probability distributions for the components of individual vortex velocities in the drift direction. Distribution functions are calculated as histograms of time averaged individual velocities in the steady state. Data are represented for different values of the reduced current $j$ close to the threshold ($j=0$). Inset: logarithmic plot of the positive tails of the above distribution. Velocity distributions are found to decay exponentially.}
\label{fig_whistov}
\end{figure}

\begin{figure}
\centering 
\epsfig{file=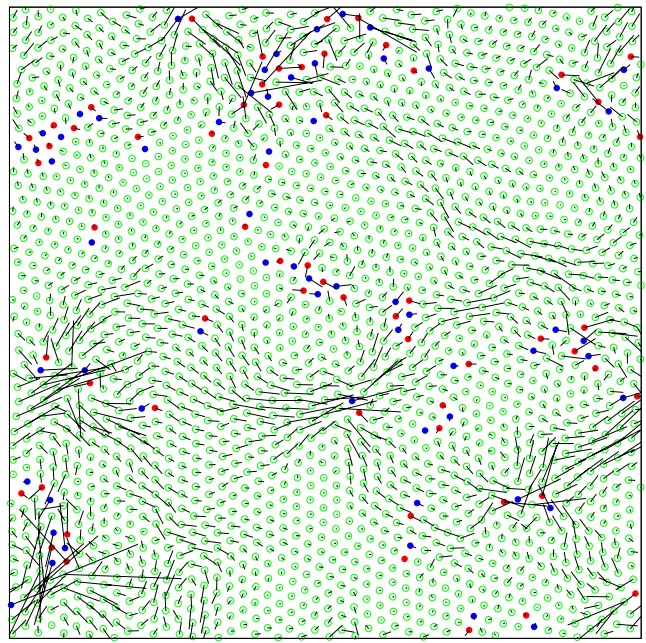,width=8.5cm,clip=}
\caption{Snapshot of the velocity field of a defected vortex lattice in the steady state, for applied currents close to the critical value. Color scheme is as in Fig. \ref{fig_wtopology}. The case of low-medium vortex densities is considered, where no healing is encountered. Also in the steady state, vortex motion is activated around moving dislocation arrays, as in the case of incipient depinning (see Reference \cite{MOR-09}). A small fraction of vortices accommodates to dislocation glide by moving in directions other than that of the applied force (from left to right in figure).}
\label{fig_wmap}
\end{figure}

The importance of dislocation dynamics in this regime can be also appreciated by looking at the velocity histograms reported in Figure \ref{fig_whistov} for the particular case of $N_v=1032$ vortices and $N_p=4128$ pinning centers. 
Probability functions for time-averaged individual vortex velocities in the drift direction at the steady state have been calculated for applied currents very close to the critical value. Compared to early numerical measurements of these quantities  \cite{FAL-96}, here we were able to consider larger vortex assemblies and currents closer to the threshold. We observe that, very close to depinning the probability distribution function has a peak almost at zero and a fastly decaying negative tail. The fact that a small number of vortices is moving {\it backwards} is a direct consequence of dislocation motion. If a lattice dislocation glides through the array during vortex flow, it produces rearrangements of the vortex positions in every spatial direction. As a result, the disordered flow of a dislocated vortex array in the direction of the applied force requires that at every instant of time a small fraction of vortices has a negative velocity in the drift direction, if a small enough force is applied.  This can be observed directly by looking at the velocity field under the given conditions, as reported in Figure \ref{fig_wmap}. 
For currents just above the threshold, vortex motion is nucleated around moving dislocation assemblies and exhibits a whirling dynamics, produced by shear stresses associated with grain boundary glide (Ref. \cite{MIS-07}). Although the average vortex velocity has a positive component along the drift direction, certain vortices have to accommodate to dislocation dynamics by moving transversely or even {\it backwards}. A similar result was obtained in Reference \cite{MOR-09}, but in the case of incipient depinning instead of the steady state. Here the low vortex densities ensure that {\it after} depinning the steady state does not produce global defect annihilation. Vortex flow proceeds, mainly along the grain boundaries, in meandering filaments. Filamentary flow is thus originated by the plastic flow of the vortex polycrystal.

\section{Strong pinning interactions: amorphous phase}\label{strong}
\begin{figure}
\subfigure[]{\epsfig{file=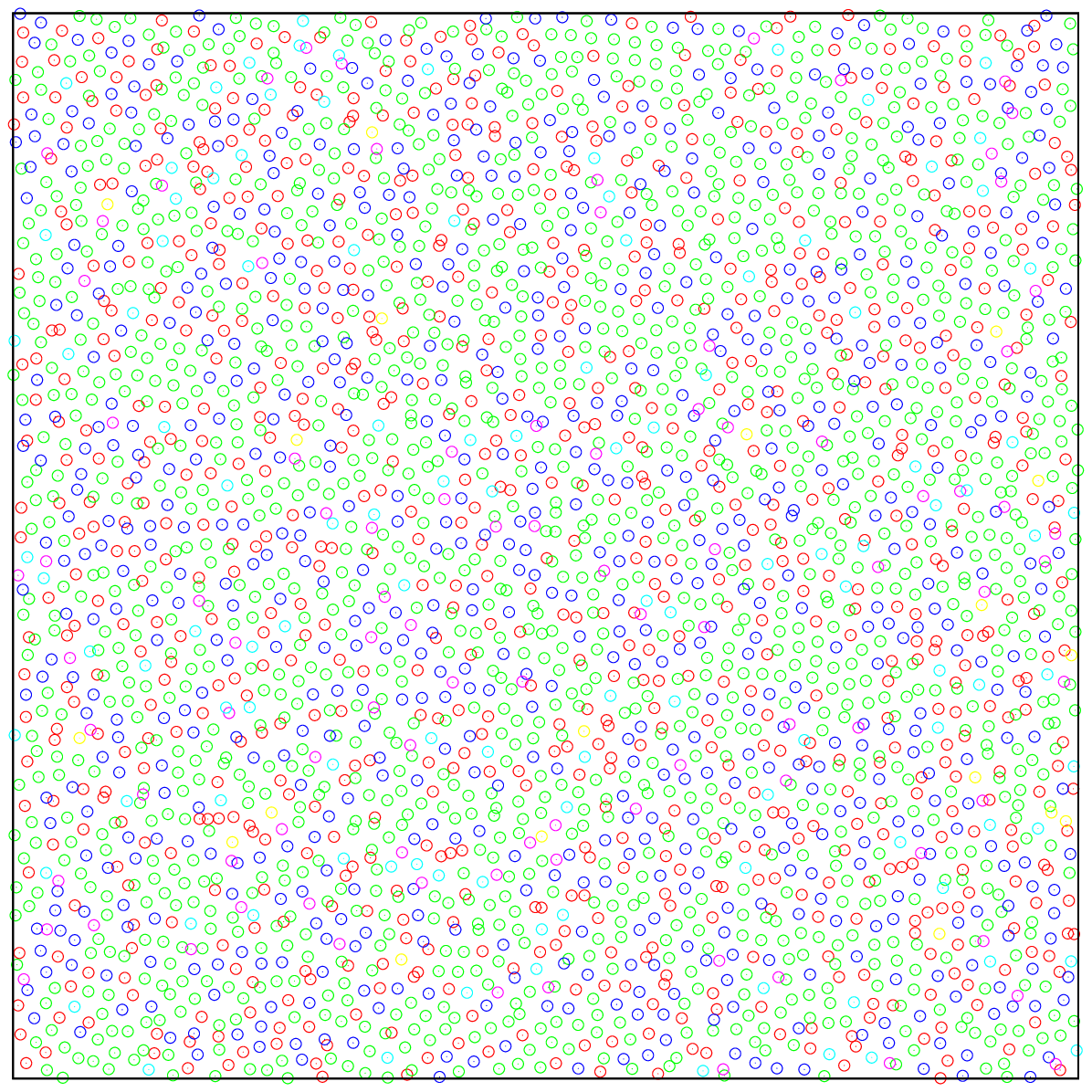,width=4cm,clip=}}
\hspace{0.2cm}\subfigure[]{\epsfig{file=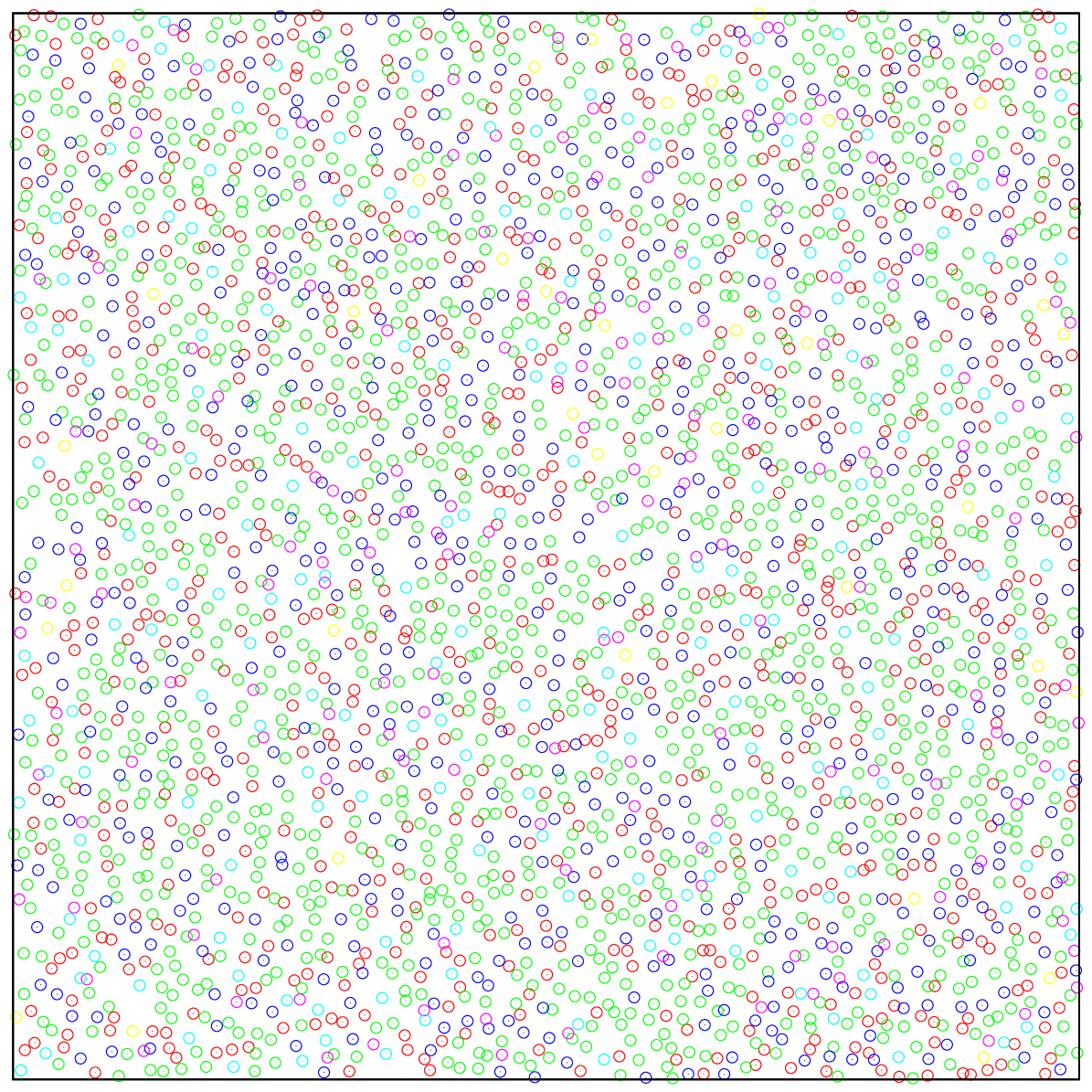,width=4cm,clip=}}
\subfigure[]{\epsfig{file=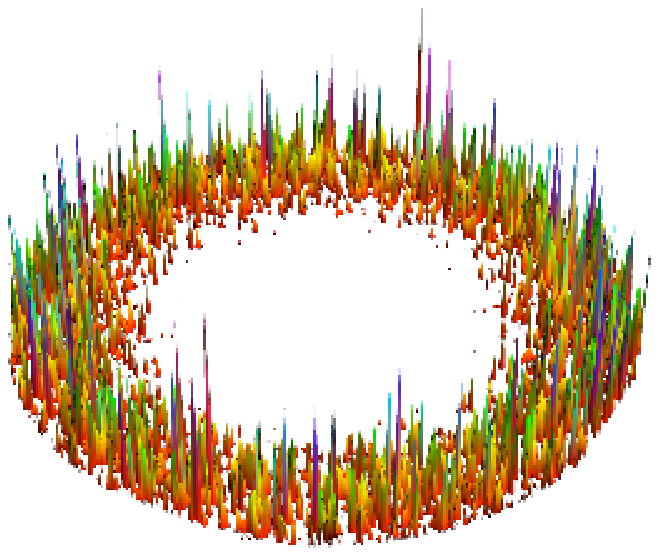,width=4cm,clip=}}
\hspace{0.2cm}\subfigure[]{\epsfig{file=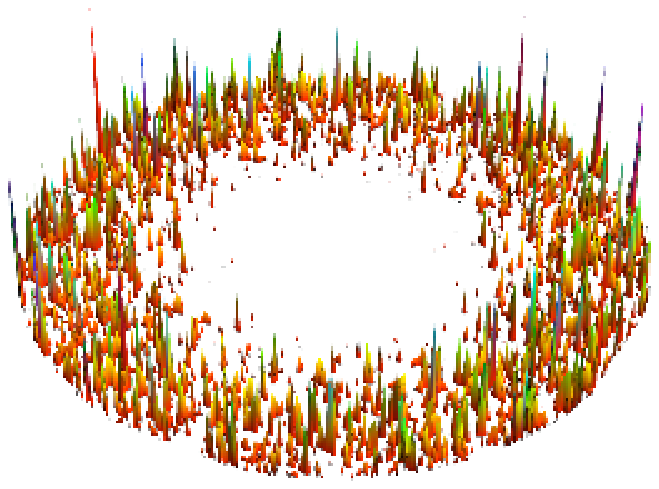,width=4cm,clip=}}
\caption{Delaunay triangulations,(a) and (b), and simulated diffraction patterns, (c) and (d),  of relaxed vortex arrays in the regime of strong pinning interactions. Data are provided for $N_v=3120$ vortices in interaction with $N_p=2064$ point impurities, (a) and (c), and $N_p=7224$ point impurities, (b) and (d). In Delaunay triangulations, each circle is an $i$-coordinated vortex, and the following scheme is applied: green for $i=6$, red for $i=5$, magenta for $i=4$, blue for $i=7$, light blue for $i=8$ and yellow for other coordination values. It appears that the dislocation representation becomes inadequate in this case, as the system is found in a purely amorphous state. To enhance visibility, vortices are drawn as thick circles, however real vortex cores are much smaller and no core overlap effect is produced. Diffraction patterns are represented with the same scale factors and cut-offs. Ring structures suggest that both translational and orientational order are lost. Upon increasing impurity density, the diffraction ring becomes thicker and less intense, meaning that the vortex spacing distribution has further broadened. }
\label{fig_stopology}
\end{figure}

We now turn our attention to the case of strong pinning interactions. In Equation \ref{pinning} we choose $V_0=1$ and perform the same study proposed in the previous section for the $V_0=0.01$ case. The first remarkable feature of this regime is that no radical difference is observed between the FC and the ZFC protocols. Both a random vortex distribution (corresponding to a high temperature phase) and a perfect vortex lattice (a {\it zero-field cooled}) phase relax in the absence of external currents into a heavily disordered vortex array, where lattice order is completely lost and the dislocation description becomes ineffective, as reported in Figure \ref{fig_stopology}. 
Such an amorphous phase should not be confused with the so-called {\it vortex glass} phase, induced by high magnetic fields and responsible for the sudden jump in the critical current known as {\it peak effect}. In this case, magnetic fields are safely low and order loss is due to the strength of impurity fields, as commonly observed in heavy-ion irradiated superconductors \cite{BEE-00}, or even in colloidal aggregates in contact with highly heterogeneous substrates  \cite{PER-08} . However one should observe that also in our case we witness an anomalous current behavior. This aspect will be investigated in the following.

\subsection{Depinning and critical currents}
\begin{figure}
\centering 
\epsfig{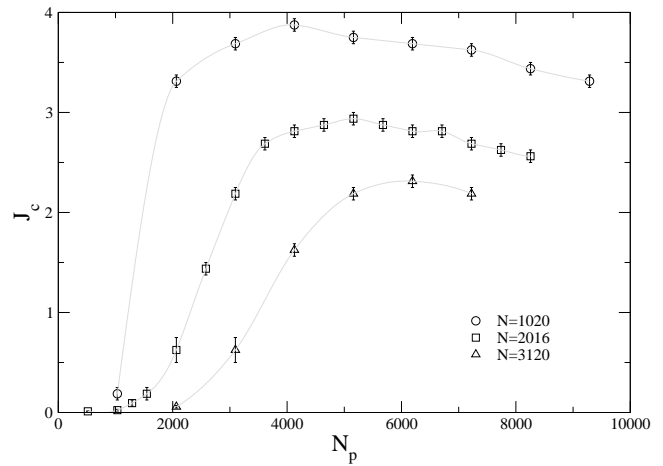}
\caption{Critical current as a function of the number of impurities for the amorphous vortex system in the case of strong pinning interactions. Three different values of $N_v$ are considered. Interpolation lines are drawn as a guide to the eye.}
\label{fig_sjcrit}
\end{figure}

As in the case of the vortex polycrystal examined above, our aim is now to investigate the behavior of the amorphous vortex array very close to the critical current $J_c$. As no high-level dislocation theory can be formulated in this case, one needs to deal with an atomistic description of the system at depinning. The simulated critical current as a function of the number of impurities is reported in Figure \ref{fig_sjcrit}. 
Critical currents are orders of magnitude higher than in the case of vortex polycrystals. In the range of $N_v$ considered there are no substantial differences between lower and higher vortex densities. The three curves start with very low $J_c$ for low impurity densities, as if the system was reminiscent of the vortex lattice, and subsequently experience a sudden jump, where the critical current increases severely and reaches a sort of plateau. In order to interpret these data correctly, we need more information about the nucleation of vortex motion at depinning.   

\begin{figure}
\centering 
\epsfig{file=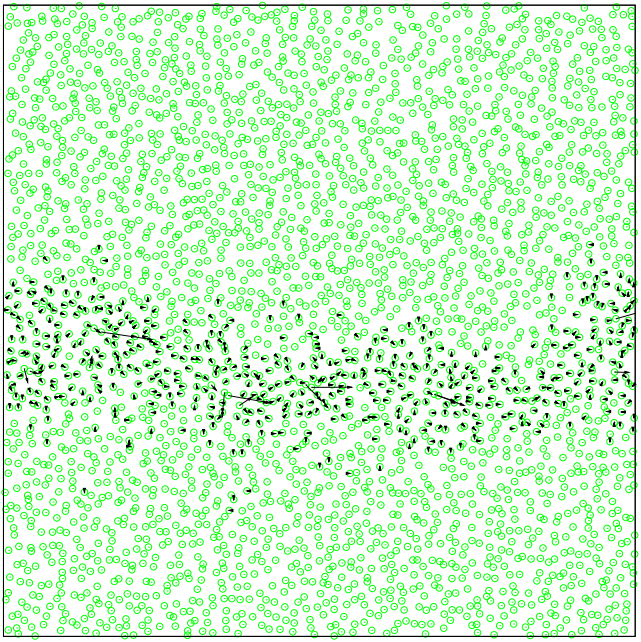,width=8.5cm,clip=}
\caption{Velocity field of the amorphous vortex array at incipient depinning, for applied currents close to the critical value. This time all vortices are represented as green circles regardless of their coordination number. Individual velocities are represented as arrows. Few vortices have a {\it macroscopic} velocity and are represented as full arrows. When velocities are very small compared to the former, but still above a certain cut-off value, they are represented as arrow heads (with no line). As velocities fall below the cut-off, no head or line is drawn. Depinning proceeds through the fast motion of few vortices along a weak direction and the slow wandering of a surrounding band.}
\label{fig_smap}
\end{figure}

In vortex polycrystals we proved that vortex activity was nucleated around depinning grain boundaries\cite{MOR-09}. The picture that emerges form our simulations is instead very different in the amorphous case. Figure \ref{fig_smap} shows the vortex velocity field at incipient depinning for the amorphous vortex array.
The system has developed a single band of activity, where a few vortices move at overly high speed roughly in the direction of the applied force (full arrows), surrounded by a stripe of vortices that are moving apparently in all possible directions (arrow heads alone), while the rest of the sample experiences velocities well below the chosen cut-off, thus appearing motionless. The interpretation of this picture is rather intuitive. In the given regime, pinning forces overpower vortex interactions, so that lattice effects are almost negligible compared to the case of a polycrystal. However, the high forces applied allow the moving vortices to come extremely close to the cores of other vortices, experiencing their strong repulsion.The external force is applied uniformly over the sample, however it is able to trigger depinning only where it finds a suitable weak spot along the force direction. Depinning is then propagated by individual vortices, a few units along the given direction. Their trajectory is rather heterogeneous, as it is the result of the interplay between the high force applied and the obstacles encountered in the way: extremely strong pinning points and other vortex cores. All around those fast  vortices, the remaining ones have to rearrange to the motion of the former, either being repelled by them or trying to fill the empty spots left behind by them. This corresponds to the band of activity in Figure \ref{fig_smap}, which becomes broader if a lower velocity cut-off is chosen. 

\begin{figure}
\centering 
\epsfig{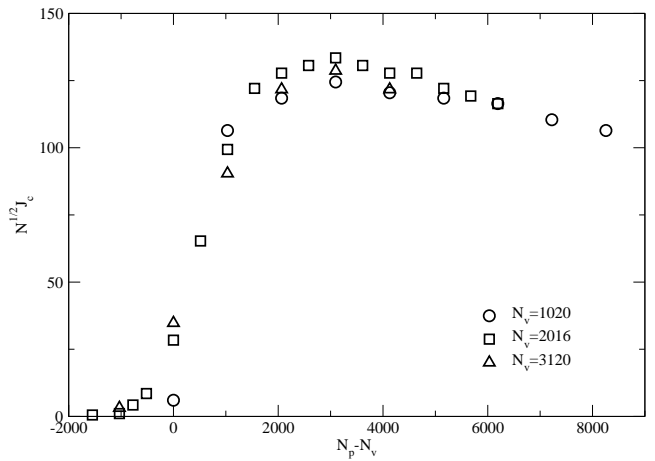}
\caption{Rescaled critical current as a function of the exceeding pinning points. Error bars were not represented in this plot. The jump in the critical current is observed when $N_p=N_v$.}
\label{fig_sjcrit_scaled}
\end{figure}

In a mean-field spirit, each fast vortex moving in the drift direction will experience interaction with pinning points and surrounding vortices as a sort of effective dynamic friction. This physical picture would then suggest that the depinning threshold should be inversely proportional to the probability $p$ of finding a weak layer in the drift direction that would yield vortex motion. Evidently $p$ is a function of the number of vortices $N_v$, as in every simulation box there are approximately $\sqrt{N_v}$ layers if disorder is isotropically distributed. Then evidently one has that to a first approximation
\begin{equation}
J_c \propto \frac{1}{\sqrt{N_v}}. 
 \end{equation}
This hypothesis is confirmed in Figure \ref{fig_sjcrit_scaled}, where the $\sqrt{N_v}$ scaling of $J_c$ is assessed.  
Figure \ref{fig_sjcrit_scaled} also shows that plotting the rescaled $J_c$ as a function of $N_p-N_v$, the three curves collapse to a master curve in which the current jump is located at the origin. The quantity $N_p-N_v$ is the number of exceeding pinning points or, alternatively, the number of {\it free} vortices changed in sign. Due to the strength of pinning interactions, vortices are individually trapped by impurities. However if $N_v$ exceeds $N_p$, at every time step there are $N_v-N_p$ vortices that are statistically free as they are repelled by {\it locked} vortices. They are natural weak spots for the system and this explains why critical currents are lower in that case. As soon as $N_p$ reaches $N_v$, instead, all vortices are locked on average, and $J_c$ increases abruptly. Any further increase in $N_p$ does not affect the system dramatically, thus explaining the final plateau region. An estimate for the critical current can thus be written as
\begin{equation}
J_c=\left[c_0\Theta(N_v-N_p)+c_1 \Theta(N_p-N_v)\right]N_v^{-1/2},
\end{equation}
where $c_0$ and $c_1$ depend on the pinning potential depth $V_0$ and are slowly varying functions of $N_p$, $\Theta$ is the step function and the magnetic field dependence is mostly conveyed by the $N_v^{1/2}$ factor. To a first approximation one has that in this regime
\begin{equation}
J_c\propto H^{-1/2}.
\end{equation}

\subsection{Steady state and coalescence}
As in the case of the vortex polycrystal examined above, we are now interested in the nature of the steady state, after depinning is produced. For the range of values considered here, we have witnessed no healing or partial recovery of lattice order. Neither vortex density nor larger applied forces seem able to reconstruct an ordered lattice. 

\begin{figure}
\centering 
\epsfig{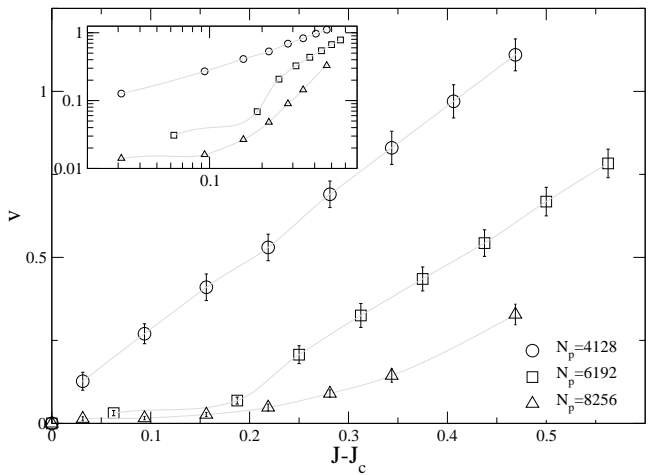}
\caption{Steady-state collective vortex velocity in the drift direction for $N_v=1020$ vortices as a function of the distance to the threshold for the amorphous vortex array. Different values of the defect density are considered. A logarithmic plot is reported in the inset.}
\label{fig_svcrit}
\end{figure}

\begin{figure}
\centering 
\epsfig{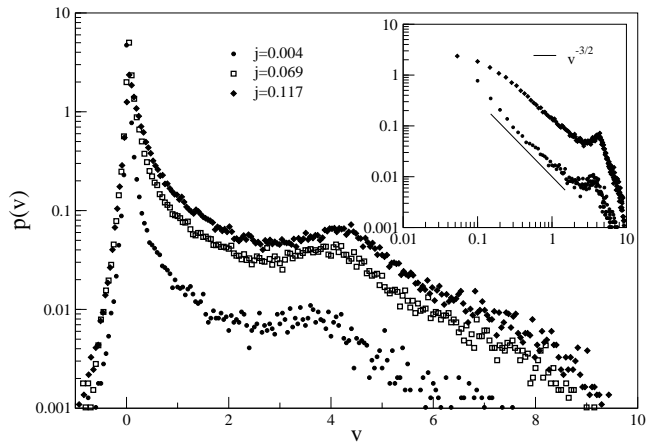}
\caption{Probability distributions for the components of individual vortex velocities in the drift direction for the amorphous vortex system. Distribution functions are calculated as histograms of time averaged individual velocities in the steady state. Data are represented for different values of the reduced current $j$ close to the threshold ($j=0$). Inset: logarithmic plot of the positive tails of the above distributions. Only two data sets are reproduced here to enhance readability. A power law with exponent $-3/2$ is plotted for comparison purposes.}
\label{fig_shistov}
\end{figure}

If the applied current is very close to the depinning value, the steady state does not differ form the picture shown in Figure \ref{fig_smap}, where few vortices are moving at high speed, while the majority are macroscopically inactive. The coexistence of a moving phase and a pinned phase can be appreciated in  Figure \ref{fig_svcrit}, where velocity-force relations are plotted for different impurity concentrations. 
The universal behavior found in the case of the polycrystal is absent here and the strongly discontinuous nature of the transition suggests coexistence phenomena, as commonly observed in experiments on vortex matter, colloids and charge density waves  \cite{JEN-88,BHA-93,PER-08,MAE-90}. A suitable way to quantify coexistence phenomena is to look at vortex velocity distributions. In Figure \ref{fig_shistov}, probability distribution functions for individual vortex velocities in the force direction are represented.    
As already pointed out in a previous numerical study  \cite{FAL-96}, such distributions are bimodal, pointing directly at the two-phase structure of the depinning system. However, in our case we looked at velocity distributions obtained extremely close to the threshold and for a larger system. This allows us to access certain quantitative aspects that could not be appreciated if such conditions were not met. Figure \ref{fig_shistov} shows two peaks. The first one is around zero velocity and quantifies the fact that most vortices are almost still. The second peak is found at very high velocities and is significantly lower, meaning that only a few vortices are moving fast. Very close to the threshold (lower data) the peak becomes up to 3 orders of magnitude lower than the peak in zero. This does not come as a surprise, because we have seen that out of thousands of vortices only few units contribute to the macroscopic dynamics if the force is just above the threshold. As we consider lower and lower currents, the tail of the peak in zero becomes clearer. The inset of Figure \ref{fig_shistov} shows that it has a power-law behavior over close to two decades in velocity, meaning that the slow phase exhibits glassy dynamics. Ideally for currents infinitesimally higher than $J_c$ and very large systems, the glassy phase would extend to the entire vortex array except an infinitesimal fraction of vortex matter. Size effects and the discrete nature of vortex matter would make these conditions difficult to achieve in experiments. However Figure \ref{fig_shistov} shows that signals of the glassy phase can also be observed at higher currents and in realistic systems.  

Compared to the case of a vortex polycrystal for weak pinning interactions, we also observe that the negative tail of velocity distributions is now a lot less prominent. As no lattice collective rearrangements due to dislocation glide are present here, vortices can temporarily move {\it backwards}, but mainly to accommodate to the motion of their fast counterparts.   

\begin{figure}
\subfigure[]{\epsfig{file=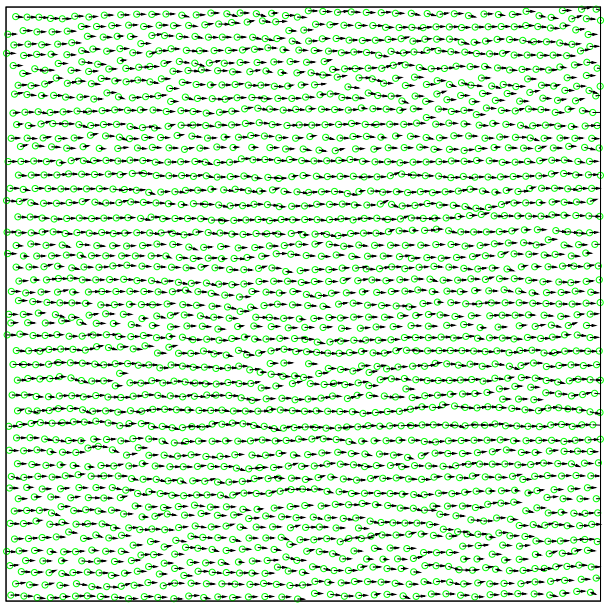,width=8cm,clip=}}\\
\hspace{0.2cm}\subfigure[]{\epsfig{file=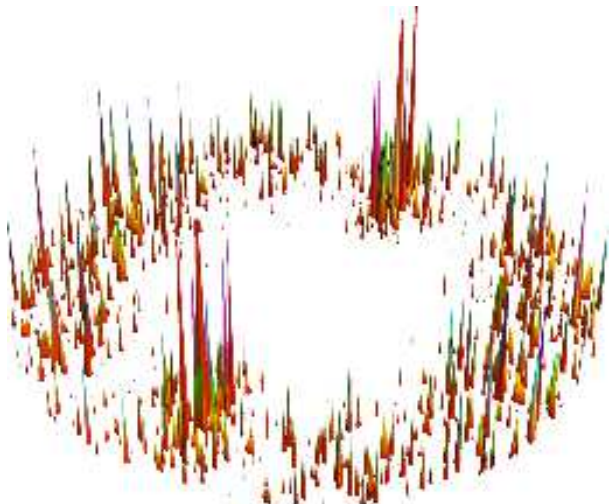,width=8cm,clip=}}
\caption{Evidence of smectic dynamics at very high drive in the case of strong pinning interactions. (a) Velocity field; (b) simulated diffraction pattern. The typical two-peak structure of the diffraction pattern accompanies smectic dynamics. Activity bands have coalesced and flow proceeds through decoupled channel motion.}
\label{fig_smec}
\end{figure}

As soon as larger currents are applied, new activity bands add to the existing one. At sufficiently large current they eventually coalesce and the entire vortex assembly flows. However, the strength of impurities is still prominent, so that the layered dynamics extends to the high drive phase, forcing vortices to move along independent channels. This is the well known smectic phase observed even in the case of weak pinning, provided that the vortex array is dilute enough. A graphical representation of the smectic phase in our peculiar regime is reported in Figure \ref{fig_smec}.

\section{Conclusions}\label{conclusions}

Topological order of driven vortex arrays in type II superconductors appears an essential ingredient to understand the flow properties of vortex matter. In most common conditions, the Abrikosov vortex lattice order is broken by the proliferation of topological defects, such as dislocations, which add new phenomenology to the intricate electromagnetic response of this physical system. In this paper we emphasize this aspect and examine two limiting cases of technological interest: i) the case of weak pinning interactions, usually due to intrinsic point disorder in the material, and ii) the case of strong pinning interactions, which can be extrinsically induced, for instance, by ion irradiation. In the first case, Delaunay triangulations and diffraction patterns unveil a polycrystalline lattice topology, with dislocations assemblies affecting the current response of the system. On the other hand, the topological analysis of the second limiting case shows a heavily disordered vortex array, where lattice order is completely lost and the dislocation description becomes inadequate.

The depinning of a vortex polycrystal is nucleated at the grain boundaries and extends to the rest of the lattice as soon as large enough currents are applied \cite{MOR-09}. Vortex flow occurs in meandering filaments originated by the plastic flow of the vortex polycrystal. We showed that {\it soft} vortex lattices are eventually unstable against the proliferation of topological defects, and both FC and ZFC protocols yield equivalent topological structures, for currents just above depinning whenever the impurity density is high enough. The average vortex velocity along the drift direction in the steady state approaches zero as the applied current decreases towards the critical current $J_c$ in a weakly discontinuous fashion.  Moreover, collective dislocation and grain boundary dynamics in this polycrystalline phase give rise to a power spectrum that decays according to a $\omega^{-1.5}$ law, in good agreement with previous results for plastically deforming crystals \cite{LAU-06}. This phenomenology is observed right above depinning in the steady state attained for low and intermediate vortex densities, where no healing is encountered and dislocation dynamics heavily affects the response of the system. In this regime, dislocation gliding also induces the backwards motion of a small fraction of vortices, which can be easily detected in the individual vortex velocity distribution.

The response of the amorphous vortex arrays resulting from the action of strong pinning interactions cannot be simply interpreted in the language of topological defects behavior. Here individual vortex behavior results crucial to determine  the depinning threshold. The amorphous vortex phase appears independently of the cooling protocol (FC or ZFC) and critical currents are much higher than for their vortex polycrystal counterparts. The magnitude of $J_c$ is inversely proportional to the probability of finding a layer along the drift direction prompt to vortex motion, and thus scales as $N_v^{-1/2}$. Moreover, it experiences an abrupt increase as soon as the number of pinning sites $N_p$ equals the number of vortices $N_v$, and thus when, on average, all vortices may be locked to a pinning point. Any further increase in $N_p$ does not alter appreciably this picture, as confirmed by the final plateau in the critical current versus pining density curve. In the steady state, the average vortex velocity along the drift direction approaches zero as the applied current decreases towards the critical current $J_c$ in a strongly discontinuous manner. Both this quantity and the bimodal distribution of individual vortex velocities evidence the coexistence of two types of vortex phases right above depinning: a glassy phase and a highly mobile phase, which sets up flow bands or channels.  

In conclusion, all the results found for these two limiting pinning scenarios emphasize the possibility of establishing a clear distinction between two dynamical and topologically disordered vortex phases: the vortex polycrystal and the amorphous vortex matter. The behavior of critical currents and their scaling properties, the velocity-force relationships, the velocity histograms and the flow geometries corroborate this viewpoint. Recent experimental findings on colloidal crystals have already illustrated the different nature of these two phases \cite{PER-08}. New experiments of a similar kind performed in vortex matter could help to ascertain the existence of a new phase boundary in the already rich dynamic phase diagram of this fascinating physical system.

\section*{Acknowledgments}
We acknowledge financial support by the Ministerio de Educaci\'on y Ciencia (Spain), under grant FIS2007-66485-C02-02. MCM also acknowledges the Generalitat de Catalunya and the Ministerio de Educaci\'on y Ciencia ({\em Programa I3}) for additional financial support. 
PM acknowledges funding form the European Social Fund through the Juan de la Cierva Programme.

\end{document}